\documentclass[%
reprint,
 amsmath,amssymb,
 aps, physrev,
prstper,
floatfix,
]{revtex4-2}

\usepackage{graphicx}
\usepackage{dcolumn}
\usepackage{bm}

\usepackage{xcolor, soul}

\begin{document}

\preprint{APS/123-QED}

\title{\textbf{High stakes exams inflate a gender gap and contribute to systematic grading errors in one introductory physics series} 
}%

\author{David J. Webb}
\affiliation{%
Department of Physics \& Astronomy,\\ University of California - Davis
}%
\author{Cassandra A. Paul}
\affiliation{%
 Physics \& Astronomy Department,
Science Education Program\\
San Jose State University 
}%


\date{\today}

\begin{abstract}
Previous research has suggested that changing the percentage of the course grade associated with exam grades in STEM courses can change the gender gap in the course. It has also been shown that assessments with the highest stakes have the lowest (relative) scores for female students. Previous research by the authors has shown that the implementation of retake exams can eliminate the gender gap in introductory physics courses. This paper explores several different hypotheses for why retake exams are associated with a zeroed gender gap. Two independent measurements comparing exams with different stakes are used in support of the argument that the entire gender gap on introductory physics exams may be due to the stakes associated with those exams. In other words, these data support the idea that a gender grade gap on exams is not measuring a gender difference in the physics knowledge or physics ability of these students. Implications suggest that instructors should choose lower stakes assessment options if they are interested in exam measurements that are not influenced by differences in students' performance related to exam stakes.
\end{abstract}

\maketitle


\section{\label{sec:Intro}Introduction}

In a recent paper \cite{webb_attributing_2023} we showed data in support of the argument that demographic grade gaps are a result of course design (the Course Deficit model \cite{Cotner2017}) rather than the more common argument that the gaps result from average deficits in specific demographic groups of students (the Student Deficit model \cite{Valencia1997}). One way we did this was by showing that the gender gap disappeared in introductory physics classes that were changed to offer exam retake opportunities to all students, while the gender gap had been persistent in regular unchanged courses. Our previous work did not examine the mechanism for how the retakes eliminated the gender gap. For example, were women more likely to take advantage of the retake opportunity, or did women improve more than men on the second exam? In this paper we analyze more detailed assessment data on the gender gap issue in these courses and argue that a reasonable interpretation of our results is that the entire gender gap in these courses is determined by the stakes associated with the exams and does not reflect a gender difference in physics knowledge, opportunity, or ability.

\subsection{\label{sec:GenGapPhysics}Gender gaps in physics}

Gender equity gaps in introductory physics courses and exams have been studied extensively over the past several decades. Gender gaps in physics can refer to performance gaps in course grades \cite{Stout2013, webb_attributing_2023}, on concept inventories \cite{Madsen2013, 10.1371/journal.pone.0271184}, and/or on exams \cite{singh_test_2021}.  Representation and retention gaps have also been observed \cite{cimpian_understanding_2020}.

\subsubsection{\label{sec:StudentDeficit}Gaps - student level characteristics}

Many different explanations for the existence of these types of gender gaps have been proposed and investigated.  Some of the leading theories for the gender gaps in physics are that women on average i) have poorer physics and/or math preparation \cite{Salehi2019, malespina_gender_2022, 10.1371/journal.pone.0271184}, ii) have less overall interest in physics \cite{sax_women_2016}, and/or iii) suffer more from test anxiety, which is amplified on high-stakes exams \cite{cotner_gender_2020, salehi_gender_2019, malespina_bioscience_2024, malespina_gender_2022, espinosa_reducing_2019, harris_can_2019, singh_test_2021}. It's also been shown that women don't feel like they belong in physics as much as their male peers \cite{CwikBelonging2022,Stout2013}, that they have less of a science or physics identity \cite{hazari_connecting_2010}, and/or have lower self-efficacy in physics \cite{malespina_gender_2022, espinosa_reducing_2019, PhysRevPhysEducRes.17.020138, PhysRevPhysEducRes.17.010143}. Finally, the context (e.g. physical vs biological or practical vs theoretical) of an exam question may influence \cite{Sencar2004,Wheeler2019} the gender gap for that question. 

In a comprehensive review of gender gaps in concept inventories, Madsen et al. \cite{Madsen2013} found that none of the 30 different factors they considered fully explained the gender gap and thus concluded that there may be many interconnected issues leading to these gaps. Despite these different competing factors, there have been a few studies that have seemed to entirely eliminate the gender gap (example: \cite{webb_attributing_2023}), suggesting that, despite the gap's persistence and the uncertainty concerning its cause, it is possible to mitigate the effects giving rise to the gap.

Exploring one potentially fruitful mitigation of gender gaps in exams, Cotner and Ballen \cite{Cotner2017} show that exam gender gaps in some biological science courses could be changed by lowering the stakes associated with exams. They compared overall success in different biology courses by categorizing them as either low-stakes (less than 40\% of the course grade determined by quizzes and exams) or high stakes (more than 40\% of the course grade determined by quizzes and exams). The men outperformed the women in the high-stakes courses, while there was no statistical difference between the genders in low-stakes courses.

Cotner et al. later suggest \cite{cotner_gender_2020} that the reason for the change in performance when the stakes change is because women are more likely to suffer from anxiety which is amplified in high-stakes situations.  In fact the connection between high-stakes exams and anxiety in women is well studied \cite{cotner_gender_2020, salehi_gender_2019, malespina_bioscience_2024, malespina_gender_2022, espinosa_reducing_2019, harris_can_2019, singh_test_2021}, where evidence suggests that women are more likely to suffer from test-anxiety that negatively impacts exam performance \cite{FrenchHighStakesCritiques2024}.  Malespina \& Singh \cite{malespina_gender_2022} found that women under-performed on high-stakes exams in calculus-based introductory physics courses. However, when they controlled for test anxiety, the gap became non-significant. There is evidence that test anxiety is an issue for women in other STEM majors as well.  Espinosa et al. \cite{espinosa_reducing_2019} found that in a large introductory biology course, women expressed higher test anxiety than men and that anxiety was a significant predictor for exam grades even when controlling for overall GPA. This is important not only in biology courses but also because a large population of introductory physics students are biology students. Similarly, Malespina et al. \cite{malespina_bioscience_2024} find the same is true for bioscience students enrolled in a physics course. Salehi et al \cite{salehi_gender_2019} study performance across three different assessment types, exams, non-exams and labs, in five different lower division STEM courses. They found that while there was a significant negative gap for women on exams, women consistently outperformed men on the other two types of assessments. Interestingly, they found that exam performance for women was negatively influenced by test anxiety in both lower and upper division courses, even though by the upper division courses the gender gap on exam scores had disappeared in both biology courses and other STEM courses.

\subsubsection{\label{sec:BrillianceIssues}Gaps - system level issues}

Leslie et al. \cite{LeslieBrilliance2015} take a more systemic view in noting that gender gaps in representation are largest in a few specific academic fields (including physics) where the practitioners in the field are more likely to think that to succeed in their field “hard work alone just won’t cut it; you need to have an innate gift or talent.”  Muradoglu et al. \cite{MuradogluBrilliancePhysics2024} use this idea, that success in a field is perceived to require innate brilliance, together with the general societal view that women are not as intellectually talented as men \cite{FURNHAM20011381}, to argue that a natural result is that, on average, women would have lower average self-efficacy in such fields.  Since, self-efficacy inversely correlates with anxiety, it also is reasonable that women, on average, would be expected to be more anxious than men during the high stakes exams in these particular academic fields.  As noted earlier, in this paper we will show data consistent with the idea that there is no gender gap in physics understanding/ability but that the commonly measured exam gender gap arises because of the stakes associated with that exam.  In this sense the exam gender gap may simply be an incorrect measurement of the underlying knowledge/ability.  It would be ironic if such illusory exam gender grade gaps were part of the evidence society used to conclude that women were not as talented as men at physics and that that conclusion led to women being more anxious on, and under-performing on, exactly these high stakes exams that led to society’s conclusions.

\subsubsection{\label{sec:CourseDeficit}Changes in course structure changes gaps}

Although the evidence for anxiety may make it a compelling explanation for the gender gap, we will continue our previous argument \cite{webb_attributing_2023} that it is the course structure that determines demographic grade gaps.  We suggest that high-stakes exams be considered the cause of the gender gap.  One reason we take this position is because high-stakes exams (being characteristics of the course structure) are easily changed, while anxiety (being a characteristic of the student) is much more difficult to change. If high stakes exams are the cause of the gender gap, changes in course structure can serve as a mitigation.

As an example for why we take this approach, consider a study performed by Simmons \& Heckler \cite{Simmons2020} where they simulated new course grades for students (after a course had already been completed) by reducing the weight of the exams in the overall course grade. They found that making the exams worth less of the overall course grade, equity gaps in the course were reduced for women and other historically marginalized students compared to their non-marginalized peers. In this example, women's anxiety was not changed, but instead the course structure was changed to successfully address the inequity. While this may seem like a semantics argument, positioning the stakes as the possible cause of the gender gap (instead of anxiety) focuses the research and intervention efforts on the changeable aspects of the course, rather than the (potentially unchangeable) aspects of entire populations. Similarly, it removes the urge to ``fix'' any perceived inadequacies among an underrepresented group (in this case, women). 

\subsection{\label{sec:RetakeIntro}Retake Exams in Physics}

If high-stakes are the cause of the gender gap, one obvious solution would be to eliminate the exams as an assessment practice. Critics of this approach say that low-stakes assessments (like homework or group projects) are not as accurate as other forms of assessment because low-stakes assessments like homework allow students to use outside resources, collaboratively solve problems and potentially even find solutions they can copy without understanding. Others claim that there is value in measuring what a student can do without outside resources, and/or in a high pressure environment, which homework assignments won't measure. While many, including Ref. \cite{FrenchHighStakesCritiques2024} and also the authors of this paper, disagree with these critiques, if exams (timed solo tests or quizzes with little to no allowed resources) are measuring something important to many instructors, eliminating them entirely is not necessarily a viable option. If so, then mitigation of the harms/biases caused by these exams might be the best one can do.

One possible way to lower the stakes of an exam, while still using exams as an assessment, is to offer a retake option for each exam given during the term.  For each in-term exam, a retake covers the same topics as the first-attempt exam, is offered after the students have seen their score on their first-attempt, is offered to every student regardless of grade, and is given a score that can supplant the first-attempt score.  Thus far, there is very little research on retake exams in large introductory STEM classes.  Davis et al. \cite{Davis_retakes_2024} find that students in a large introductory human physiology course who took advantage of retake exams on average increase their grade by 6\%. Kortemeyer et al. \cite{kortemeyer_retaking_2006} share the benefits of retakes as learning tools in introductory physics. However, none of the examples we found examined the relationship between the retakes and the gender gap in physics courses aside from our own previous work on this same dataset.

\subsection{\label{sec:EquityModels}Models of achievement and assessment equity}

This work, like our prior research \cite{webb_attributing_2023} utilizes two theoretical models. We define a course as being equitable if student grades are not predictable by demographic, meaning that all demographic groups achieve the same overall grades on average. This is the definition of an Equity of Parity model of equity \cite{Rodriguez2012}. We also use Cotner \& Ballen's \cite{Cotner2017} Course Deficit Model. A Course Deficit Model assumes that any gaps in achievement within the course between any demographic groups are a product of the course structure. Importantly, this does not mean that inequities prior to the course do not exist, the course deficit model merely assumes that the equity gaps measured within the course are a product of the course itself. Our previous work \cite{webb_attributing_2023} shows that by changing course structures it is possible to eliminate equity gaps and thus achieve equity of parity without changing the course content or the intellectual level of the content. 

We have previously studied grading and assessment in this same physics course and have found that percent scale grading provides an additional penalty to underrepresented minority groups when compared to 4.0 scale grading \cite{Webb2020, Paul2022}.  We also find \cite{Paul2018} that on average, women leave slightly more blanks on exams than men, while men are more likely to miss entire quizzes. Given that there seem to be both structural biases and behavioral test-taking differences between different demographic groups, it seems prudent to examine the impact of certain assessment strategies on different demographics. 
  
\section{\label{sec:Questions}Research questions}

In this paper we make the assumption that because the gender gap decreased or was entirely eliminated in courses that use a retake exam option, that the reason for increased gender gaps in the intro physics courses in our data set is likely due to the assessment strategy in courses. This leads us to a few different research questions:

\begin{enumerate}
  \item Does the gender gap decrease, in course sections that allow retakes on exams, because women students are more likely to take advantage of the retake opportunity?
  \item Does the gender gap decrease, in course sections that allow retakes on exams, because women students are doing better on their second try? 
  \item How are the stakes of an exam correlated with the gender gap? 
\end{enumerate}

Each research question tests an assumption that either came from the literature or was proposed to the authors of this paper in conversations about our previous work \cite{webb_attributing_2023}.  The first research question examines the assumption that women and men have behavioral differences that influence how likely they are to take advantage of optional opportunities to improve their grade. The second research question tests the assumption that because women are less likely to be prepared to take physics, that they might need more time to practice and receive help before succeeding than men do. The third research question was born out of negative results from the first two questions.

\section{\label{sec:Data}Database Details}

All of the exam data we discuss here are from classes in the introductory physics series taken mostly by bioscience students and offered between 1997 and 2015 at an R1 public university.  These dates were chosen because these years included the retake classes, and because the classes that allowed students to take retake exams are part of the same course series as those that did not allow retakes.  The topics covered in the course series did not change over the set of years included in the study and most of the specific student activities are also essentially the same over those years.  We chose this set of years because we had access to individual exam scores for almost all of those years but for no later years.  We have access to individual exam scores (both first-take and retake scores) for three of the four classes that offered students retake exams.  In our previous paper \cite{webb_attributing_2023} we agued that the retake effect appeared to be independent of instructor because these instructors also taught non-retake classes and their course grades showed clear gender gaps in those classes.  Finally, exams in these courses are open-ended in the sense that the students are expected to provide both a written discussion of some particular problem as well as calculate some result(s).  All of the exams given in the retake courses were drawn from a pool of exams that had been given previously in these courses so we consider them to be representative of usual exams in these classes.

To compare the exam scores from the retake classes to the exam scores from the regular classes which did not offer retakes we add, to the database, exam data of the 254 non-retake classes from the same series for which we have scores from exams given to students during the term.  We also added each student's gender and race/ethnicity as well as each student's university GPA upon entering the class.  These data were obtained from the university administration and included before anonymizing the entire dataset several years ago.  Since the numerical exam scores are given in a wide variety of ways by the many instructors of these classes, we have z-scored each exam in each course to give each exam grade distribution the same average, 0, and the same standard deviation, 1.

All together the database included 26,783 students with 16,493 identified as female students, 10,254 identified as male students, 5 identified as non-binary, and 31 students with unknown gender identities.  This university started collecting self-identified gender information in the Fall term of 2015 so our dataset is almost completely binary gender only.  For completeness, the data regarding the small numbers of unknown gender and non-binary students are discussed in Appendix \ref{sec:Nonbinary}.  The exam data include 5,598 students appearing only once, 7,382 students appearing in only two of the classes and only once in each, and 13,268 students appearing exactly once in each of the three classes in the 3-course series.  The three retake classes had a total of 397 students.

The number of individual exams given during the term varied from a minimum of 2 midterms given in each of seven classes to a maximum of 9 weekly exams given in each of 6 classes with the most common number given being 4 in-term exams in each of 115 classes (three of those are the retake classes).  In addition, the databases from 203 classes also had final exam grades and so those classes had all exams accounted for.

\subsection{\label{sec:Stakes} Classifying the Stakes of an Assessment}

For the purposes of this paper we will classify the stakes of an exam into three different categories. We use the same definition of high-stakes assessments as those typically used in the literature \cite{Cotner2017}. ``High-stakes'' assessments are timed exams that are usually in-person and allow limited use of resources. For example, a student might be able to bring a note card, but they would not be able to use the internet during the exam. In addition, these exams are considered high-stakes because their scores are a large factor in determining the course grade.  This is an implied distinction in previous work because high-stakes exams are often differentiated from quizzes that are also usually timed and allow limited resources \cite{Cotner2017} but have scores that are not as important factors in the course grade.

Instead of examining stakes of different assessment types (e.g. homework vs. exams) as other have done, in this paper we are investigating if the differences in the stakes of different exams have an impact on equity gaps found in those exams. To that end, we will take the stakes of any particular exam to be a measure of the importance of that exam's (or quiz's) score to the course grade. For instance, in the classes included in our database, the grade on the final exam constitutes roughly 50\% of the course grade (no courses in our dataset allowed retakes on final exams). Therefore, the final exams in either type of course are considered ``high-stakes" as compared to other exams. 

We also introduce the idea of a medium stakes assessment. In non-retake courses, the entire set of two to nine in-term exams (in-term meaning excluding the final) constitute roughly the other 50\% of the course grade so each of those exams has considerably lower stakes than the lone final exam. We consider these exams given as the term progresses to be ``medium stakes'' both because each individual in-term exam counts for a much smaller portion of the course grade than the final exam and because usually the class syllabus described in-course grading rules telling students that the lowest in-term exam would be dropped before calculating the course grade. 

On the other hand, in retake courses, the entire set of four first-attempt in-term exam scores \textbf{may have no contribution at all to the course grade} because the retake exam score is always available to supplant the first-attempt exam score. For this reason we consider a first-attempt exam in a retake course to have relatively lower stakes than any of the in-term exams in a non-retake class, and therefore define the first attempt on a retake as a ``low-stakes'' assessment.

We consider the first try on an in-term exam that has a retake option to be a low-stakes assessment because a second try is free to any student and could completely erase that first score. But we consider the second attempt in a retake class to be a medium stakes assessment because this grade will contribute to the final grade at roughly the same weight of the in-term exams in the courses that do not have retake exams.  All of these definitions of the stakes of different exams are outlined in Table \ref{tab:tabStakes}.

Note that the medium-stakes exams in the retake classes (the actual retake exams) are only taken by a subset of students who choose to take advantage of the opportunity to do so. This will make it difficult to draw conclusions from this data set so we will leave any calculations of gender gaps within these student subsets, varying through the term, to Appendix \ref{sec:RetakeScores}.

\begin{table}[htb]
\caption{Descriptions of the magnitude of the stakes associated with an exam.  The stakes associated with one exam can be thought of as determined by the minimum amount that that set of exams can possibly contribute to the course grade divided by the total number of exams of that type.  Retake courses and regular courses each had a single final exam so all final exams had high stakes.}
\label{tab:tabStakes}
\begin{ruledtabular}
\begin{tabular}{ c  c  c }
& & \textbf{Min. possible} \\
\textbf{Exam} & \textbf{Exam} & \textbf{contribution } \\
\textbf{Stakes} & \textbf{Type} & \textbf{to course} \\
& & \textbf{grade by an} \\
& & \textbf{average exam} \\ 
\hline
 &\textbf{Retake class} - Set of & \\
Low & four first-attempts & 0\% \\
& at short in-term exams & \\
\hline
 & \textbf{Non-retake class} - Set & \\
& of four to nine & \\
& short in-term exams & \\
Medium & or two midterm exams & 5\%-15\% \\
& \textbf{Retake class} - Second- & \\
&  attempt at in-term & \\
& exams (i.e. the retakes) & \\
\hline
 & \textbf{Both types of class} - & \\
High & One long final exam& 40\% \\
\end{tabular}
\end{ruledtabular}
\end{table}
\section{\label{sec:Results}Results}

\subsection{\label{sec:LookAtRetakes}In-term exams in retake classes}

In a previous paper \cite{webb_attributing_2023} we showed that the gender gap in course grades was zero for a set of four classes that offered retake exams to their students.  In a retake class students took an in-term exam, got their scores on that exam, and then (the next week) had a chance to take a different exam (the retake exam) that covered the same material.  Students were told that if they scored higher on the retake then the retake score would supplant the grade on their first attempt but if they scored lower on the retake then the retake and the first attempt scores would be averaged to give the recorded exam score. In these classes, there were a total of four in-term exams and so four in-term retake exams. Here we take a closer look at the details of these first attempts and the exam scores after the subsequent retakes.

Our expectation was that students who had high scores on their first attempt would not risk lowering that score and so would not take a retake exam. We found that on average a first attempt of an exam was followed by 34\% of the students taking the retake a week later. The retake classes were anomalous in having zero gender gap in course grades (i.e. female students had the same average course grades as male students unlike in the regular courses) suggesting the following possible gap-closing scenario regarding retakes: i) female students may be more likely to take retake exams to improve their grade and/or ii) female students may make larger gains on retake exams.  Examination of the data shows that the first possibility is wrong.  We find that 78\% $\pm$ 4\% of male students retook at least one exam and 74\% $\pm$ 3\% of female students retook at least one exam (a $\chi^2$ test showed $\chi^2(1, N=399)=0.89$ with $p=0.35$ so no measurable gender difference). All together male students retook an average of 1.41 $\pm$ 0.09 exams and female students retook an average of 1.33 $\pm$ 0.07 exams (a t-test gives $p=0.49$ so no difference here either). It seems that the two genders had about the same need for retake exams.

We can examine the second possibility mentioned above by measuring the gender gap on each exam and also the gender gap after each retake had been scored.  As noted in section \ref{sec:Data}, the grades from the 12 exams from the three different courses were normalized so that i) each set of first-attempt scores from each exam in each course has the same average, = 0, and standard deviation, = 1, and ii) the recorded grades (i.e. after retake) from each exam from each class were similarly z-scored.  The result of this is that each gender gap will be measured in units of standard deviations of the exam scores for that exam.  We will also control for each student's incoming GPA in our estimates of each in-term exam gender gap.  We do this because the exam-score gender gap in a class has a small but clear dependence on the GPA gender gap in that class (see Appendix \ref{sec:RetakeExamResults} which shows these class-dependent gender gap data) and we wish to remove that dependence from our in-term exam results.  We control for GPA using hierarchical linear modeling (HLM) with the students as lowest level in the hierarchy, the set of classes as the next higher level, and then the part of the 3-course series (A, B, or C) makes up highest level of the hierarchy.  We use all three levels because our retake classes cover all three levels.  With HLM the data are fit first class by class and then those fits are combined into A, B, and C results which are then combined to give the final coefficients and their errors.  We use HLM to account for class-to-class differences in exams and A, B, and C differences in topics covered since any of these may be affecting the gender gaps and we want the final error estimates to reflect all of these variations.  For more detailed information on HLM see reference \cite{VanDusen2019}.  We use the variables $GPA$ for a student's GPA and $Female$ = 1 for female students and = 0 for male students and fit the following model eight times, once for each of the four first-attempt exam scores taken through the term and once for each of the four exam scores recorded after the retake scores are factored in,
\begin{equation}
InTrmExm = b_0 + b_{GPA}GPA + b_{Female}Female
\label{eqn:IndivQGenderGapModel}
\end{equation}
so $b_{Female}$ is the (GPA-controlled) gender gap for each fit of in-term exam scores.  We also average each student's first-attempt exam scores over the term and fit our model to those and we average each student's after-retake scores and fit our model to those.  So $InTrmExm$ can be one of the four first-attempt exam results, one of the four after-retake exam results, or one of the averages over the full term.

\begin{figure*} [htbp]
\includegraphics[trim=1.3cm 4.0cm 3.6cm 4.7cm, clip=true,width=\linewidth]{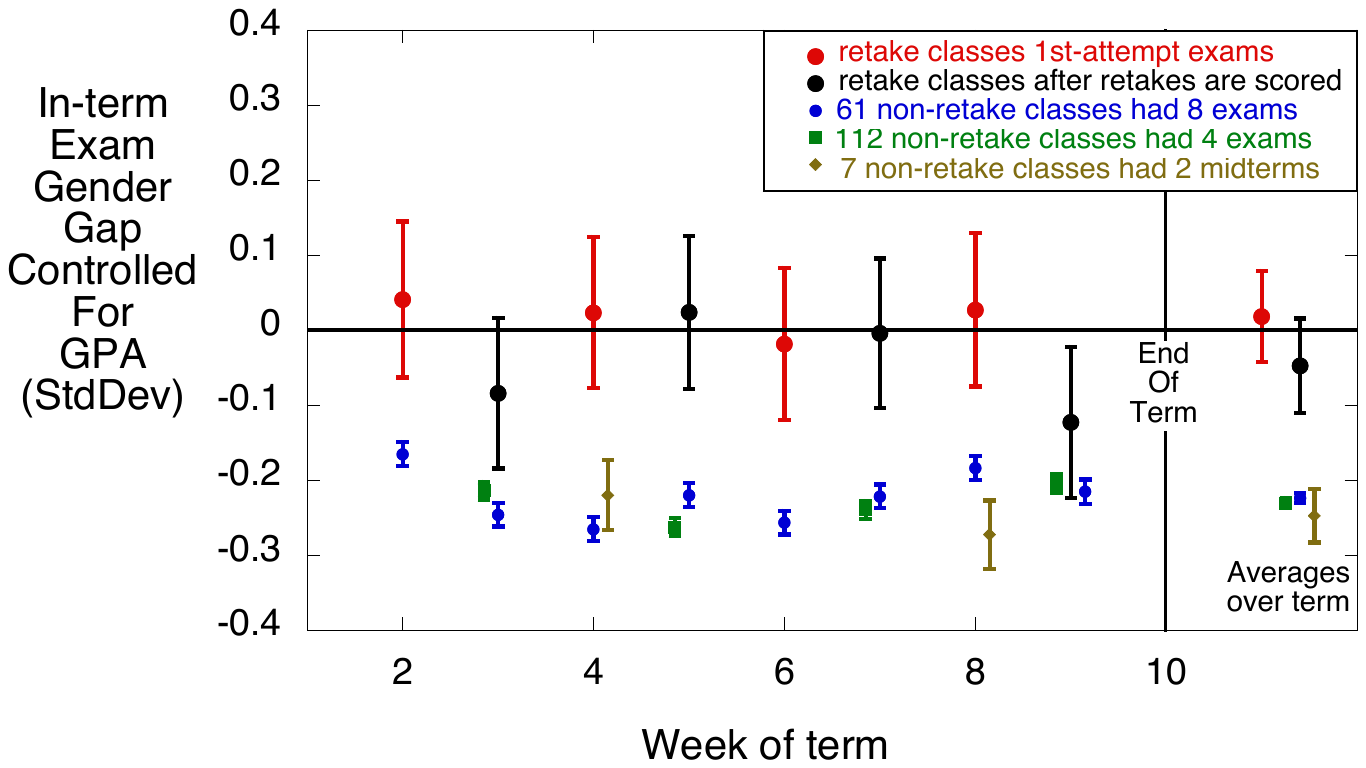}
\caption{Each in-term exam was normalized on a class-by-class basis so that the standard deviation of the grade distribution is one.  The gender gap for each group of classes was determined with Hierarchical Linear Modeling while controlling for each student's incoming GPA.  This in-term exam score gender gap is plotted as a function of time through the 10-week term at the times the exams and retakes were taken.  On the right, after the end of the term, are the overall averages for those first-attempt exams and for the recorded scores after retakes were completed.  One sees that the female students generally had higher exam scores than male students (i.e. a positive gender gap) after the first attempt and that, overall, retakes favored the male students.  For reference the gender gaps for 180 non-retake classes are also shown on a week-by-week basis through their terms.}
\label{fig:GenderGapsThruTheTerm}
\end{figure*}

Figure \ref{fig:GenderGapsThruTheTerm} shows these gender gaps for each first attempt and also after each retake exam.  These results are plotted as a function of the week the exam was taken through the 10-week term.  The overall averages, after all four in-term first-attempts and their retakes were graded, are shown to the right of the end of the term.  Notice that \textbf{female students outscored male students on three of the four first-attempt in-term exams}.  And the average of all first-attempt exams is slightly positive (female students out-performing male students) but the average of all in-term exams after retakes are scored is slightly negative (retake exam scores benefited male students slightly more than female students).  So the idea that the retakes allowed female students to improve their first attempt scores in a way that reduced the gender gap is also untrue.  This result suggests a third possible effect of allowing retake exams.  iii) The knowledge that retakes are allowed lowers the stakes of each first-attempt from what it would be without retakes.  So it may be that the scores are gender neutral because the stakes of the first-attempt exams are low.  Specifically, we suggest that the first-attempt has lower stakes in a retake class than it does in the regular nonretake classes where the first attempt is also the last attempt. In Figure \ref{fig:GenderGapsThruTheTerm} we have also included the week-by-week GPA-controlled in-term exam gender gaps for three sets of non-retake classes for which the week-by-week exam pattern through the term is known. In the rest of the paper we use the ideas of low-stakes, medium-stakes, and high-stakes exams in comparing the retake classes with the 254 non-retake classes for which we have exam data.

\subsection{\label{sec:InTermExams}Low-stakes versus medium-stakes assessments}

Instructors in this series of classes determined course grades almost completely by averaging the exam grades using a weighting of their choice. A weighting that was common (almost universal) was to 1) drop the lowest score among the in-term exams (either one in-term exam or one-half of one midterm) and then 2) weigh the final exam somewhere between 40\% and 60\% depending on whether the in-term exam average was higher or lower, respectively, than the final exam score. The most common numbers of in-term exams were four and eight so the grade weight of a single in-term exam ranged from roughly 5\% to 15\% of the course grade. We consider these exams to be medium-stakes assessments. The final exam carries several times the weight of any particular in-term exam so we consider it to be a high stakes assessment.  For the retake courses we consider the first-attempt of every exam set that students take to have low stakes because the student knows that they are free to take another exam and discount the first one if they choose to. In all classes (retake and non-retake) it was not possible to retake the final exam.

We are interested in comparing the gender gap resulting from low-stakes in-term exams with that from medium-stakes in-term exams. For this reason we compare the gender gap in the \textbf{first-attempt exams} from the retake classes with the gender gaps in the 254 non-retake classes in our dataset. Since different classes had different numbers of in-term exams we will just average the z-scored in-term-exam scores for the entire term for each student in each class and model the term-averaged in-term exam z-score for each student. Also remember that the gender gap for in-term exams in a class has a small but clear dependence on the gender gap for GPAs in that class so we will control for GPA in our analyses.

We model the comparison of the 3 retake classes with the 254 regular classes using the same kind of 3-level hierarchical regression analysis that was previously discussed. Again, the student level is the lowest level, then students are grouped in classes (2nd level), and the classes are grouped within the course series (A, B, and C in the series identify the highest level).  We use all three levels because our database covers all three levels and also because the different classes may have slightly different grade penalties for female students. We again use the variables $GPA$ for a student's GPA, $Female$ = 1 for female students and = 0 for male students, and now include $Retake$ = 1 for the three retake classes and = 0 for the 254 regular classes.  The model we fit for the student's in-term exam z-score is 
\begin{equation}
\begin{split}
InTrmExm & = b_0 + b_{GPA}GPA \\ &+ b_{Female}Female + b_{Retake}Retake \\ & + b_{Female*Retake}Female*Retake
\end{split}
\label{eqn:QGenderGapModel}
\end{equation}
where the coefficient $b_{Female}$ represents the GPA-controlled gender gap and the coefficient $b_{Female*Retake}$ measures the change in the gender gap when retakes are possible.

The coefficients resulting from this HLM fit are given in Table \ref{tab:InTermExamFit}. Recall that we consider the first-attempt exams we are using for the retake classes to be low stakes.  In the 254 non-retake classes which had medium-stakes in-term exams the average grade penalty ($b_{Female}$) that female students receive over the term is about 0.23 standard deviations (of the exam average distribution). The interaction coefficient ($b_{Female*Retake}$) for classes with low-stakes exams also has magnitude 0.23 but is positive so that the low-stakes exams net gender gap is, as expected, approximately zero. Seeing a gender gap disappear when the only change seems to be the stakes of the exams might cause one to pause before claiming that the gender gaps in the medium-stakes exams are actually measuring a gap in physics knowledge/ability.

Note that even though the coefficient for $Retake$ is negative, indicating (perhaps) that men do worse under retake conditions, this is only because the grades are z-scored. In our previous paper \cite{webb_attributing_2023} we showed that, on average, both genders had higher grades in the retake classes.  Also, as shown in Appendix \ref{sec:MoreEthnicities}, the gender gap decreases in size in retake classes for each racial/ethnic group which are sufficiently represented in our data.  Finally, as discussed in Appendix \ref{sec:RandStudInclusion}, to decide if these coefficients or our resulting conclusions are affected by the fact that many students are in our database more than once (because they have taken more than one of the classes in this three-term series) we have tested the effects of limiting each student to just one appearance in the database.  We find that the coefficients and our resulting conclusions are unchanged.

\begin{table}[htbp]
\caption{The coefficients from an HLM fit to Eq. \ref{eqn:QGenderGapModel} are shown along with their standard errors, z-statistics, and p-values.  Included are 257 classes that had a total of 57,806 students whose GPA's are known.  The interaction term suggests that the gender gap is significantly different (reduced) for the retake classes.}
\label{tab:InTermExamFit}
\begin{ruledtabular}
\begin{tabular}{c c c c c}
\textbf{Coeff.} & \textbf{Value} &\textbf{Error} & \textbf{z-statistic}
& \textbf{p-value}\\ 
\hline
$b_{GPA}$ & 0.8121 & 0.0047 & 173.0 & $<10^{-3}$ \\
$b_{Female}$ & -0.2262 & 0.0046 & -49.0 & $<10^{-3}$ \\
$b_{Retake}$ & -0.169 & 0.052 & -3.24 & 0.001 \\
$b_{Female*Retake}$ & 0.234 & 0.057 & 4.10 & $<10^{-3}$ \\
$b_{0}$ & -2.310 & 0.018 & -128.1 & $<10^{-3}$ \\
\end{tabular}
\end{ruledtabular}
\end{table}

\subsection{\label{sec:FinalExams}Testing the stakes hypothesis using the highest stakes assessments}

Since 203 of the classes in our database have final exam scores it seems useful to examine those data because the final exam scores contribute much more to the course grade than any in-term exam and so can be considered to be high-stakes exams. Unfortunately we don't have a low-stakes version of the final exam to compare with the high-stakes versions because even in the courses with retake options on the exams, it was not possible to retake the final exam. However, we do have the medium-stakes in-term exam scores in each class which should be good predictors of final exam scores in that class because they are written and graded by the same instructors and largely cover the same material. In fact, we find a correlation coefficient between the in-term exam average and the final exam score of 0.65 (with $p<10^{-3}$) but a smaller correlation between student GPA and either in-term or final exam score (both correlation coefficients are about 0.56, $p<10^{-3}$).  We've already suggested that the medium-stakes in-term exam scores include a bias against female students, so it may be that using these scores as predictors of final exam scores will not only control for the demonstrable knowledge/ability of each student in the particular course but may also control for any gender bias inherent in exams. If this were the case then the controlled final exam gender gap should be approximately zero. To test this idea we will use HLM to fit the following model
\begin{equation}
\begin{split}
FinalExam & = b_0 + b_{InTrmExm}InTrmExm \\ &+ b_{Female}Female + b_{Retake}Retake \\ & + b_{Female*Retake}Female*Retake
\end{split}
\label{eqn:FinalGenderGapModel}
\end{equation}
where we include the variable $Retake$ and the interaction term even though the retake courses had the same sort of high-stakes final exam as all the other courses.  If the gender bias on in-term exams is roughly the same as the gender bias on final exams then the coefficient $b_{Female}$ from this fit should be nearly zero.

The resulting coefficients from the HLM fit to Eq. \ref{eqn:FinalGenderGapModel} are given in Table \ref{tab:FinalExamFit}.  The most notable coefficient is $b_{Female}$ and the most notable thing about this coefficient is that it is quite clearly not zero. In fact, the (in-term exam controlled) final exam gender gap is half the size of the gender gap for the in-term exams alone.  This tells us that the full gender gap on the high-stakes final exam is about 50\% larger than that of the medium-stakes in-term exams. So the high-stakes final exam has a much larger gender gap than the medium-stakes in-term exams which, themselves, have a much larger gender gap than the low-stakes first-attempt in-term exams in the retake classes.  And $b_{Female*Retake}$ is consistent with there being a similar (in-term-exam controlled) gender gap in the high-stakes final exam in the retake classes. Again, it seems likely that the importance (or weight) of the exam (the exam stakes) is closely related to the size of the grade penalty received by female students.  Finally, as noted in Appendix \ref{sec:RandStudInclusion}, we have checked that these coefficients and the resulting conclusions hold when we (randomly) include each student only once in our analyses. Again, as in section \ref{sec:InTermExams}, it seems that exams with higher stakes have higher gender gaps.

\begin{table}[htbp]
\caption{The coefficients from an HLM fit to final exam z-scores (equation \ref{eqn:FinalGenderGapModel}) are shown along with their standard errors, z-statistics, and p-values.  Included are 203 classes that had a total of 48,428 students.  The interaction term suggests that the gender gap is not significantly different for the retake classes.}
\label{tab:FinalExamFit}
\begin{ruledtabular}
\begin{tabular}{c c c c c}
\textbf{Coeff.} & \textbf{Value} &\textbf{Error} & \textbf{z-statistic}
& \textbf{p-value}\\ 
\hline
$b_{InTrmExm}$ & 0.9656 & 0.0052 & 187.4 & $<10^{-3}$ \\
$b_{Female}$ & -0.1155 & 0.0071 & -16.2 & $<10^{-3}$ \\
$b_{Retake}$ & 0.025 & 0.066 & 0.38 & 0.705 \\
$b_{Female*Retake}$ & -0.034 & 0.080 & -0.42 & 0.676 \\
$b_{0}$ & 0.0744 & 0.0056 & 13.23 & $<10^{-3}$ \\
\end{tabular}
\end{ruledtabular}
\end{table}

\section{\label{sec:Discussion}Discussion}

In our introduction we present many student-level variables that may contribute to the gender gap in introductory physics. But these factors; preparation, interest, anxiety, stereotypy threat, self-efficacy, identity, and belonging are all deeply intertwined with both larger influences of society, and the cultural context of the physics discipline. For instance, results from Leslie et al. \cite{LeslieBrilliance2015} on the gender imbalance in academic fields (including physics) that are felt to require ``innate talent'' rather than persistent effort has led these researchers as well as others (Muradoglu et. al. \cite{MuradogluBrilliancePhysics2024}) to argue that, in fields perceived to require ``innate talent'' women (and other groups that are often intellectually marginalized) have more negative experiences, feel less welcome, have more ``imposter'' feelings, and are more vulnerable to stereotype threat \cite{Dar-NimrodStereotypeThreat2006}. This general idea can connect the various gender gap explanations that we have listed above. We postulate that many of these feelings of otherness are exaggerated in a high-pressure environment. With this stated assumption we can then argue that the reason the exam gender gap exists is not because of all these negative biases against women, it's because these biases exist AND we put students in a high pressure environment. 

Furthermore, an extremely important takeaway from this logic is that tests are not measuring what we think they are measuring because if they were, performance gaps between demographic groups would not depend on the stakes of the particular exams as Sections \ref{sec:InTermExams} and \ref{sec:FinalExams} show they do. These two sections provide two independent results, each supporting the idea that increasing the stakes of an exam increases the gender gap in a way that is disconnected from differences in physics knowledge/skills.  The implications are that if we care about measuring physics knowledge and skills, we need to be looking at ways to reduce or eliminate the stakes associated with our assessments.  Since the stakes of a given timed in-person assessment correlate with the size of the gender gap, exams are currently not good measures of physics knowledge and skills across demographic groups. 

Another problem with putting one's focus on the individual student level issues that large groups of students may be having in physics is that this leads too naturally to a consideration of those groups of students as deficient or inadequate in some way. Researchers tend to use a Student Deficit model rather than the idea that these students are just reacting, on average, the way anyone would if they were part of a group stereotyped by society. Indeed, results from Salehi et al \cite{salehi_gender_2019} fit this picture - they find that the gender gaps for exams are much larger in general science and engineering courses where women are less represented compared to biological science courses where women are more represented and may therefore feel more like they belong in that space. 

Cotner and Ballen \cite{Cotner2017} cut the Gordian knot made up of interconnected possible student deficits by suggesting the easiest thing to do is to simply change the structure of the course. They proposed using a Course Deficit model rather than a Student Deficit model. Our recent work \cite{webb_attributing_2023,paul_examining_2025} supports the idea that changing the course structure is enough to eliminate equity gaps. While studies that examine the effects that society (and, more specifically, disciplinary cultures) have on different demographic groups are important and necessary, we should not present those effects as the reasons for the differences between the performances of demographic groups. If we move in that direction we miss the forest for the trees. Perhaps the most important take-away from our paper is that the solution to decreasing the gender gap in physics exams isn't to make women feel less anxious in a stressful environment, it's to change the environment to be less stressful. This is the power that we have as instructors. We might not be able to change the impact society has on women in physics, but we can mitigate the effects of those impacts in our courses.
The implications of this work indicate that instructors should strongly consider adopting assessments such as  retakes that lower the stakes associated with testing in introductory physics courses. Timed in-person exams are typically anxiety inducing experiences, however our work shows that if we change the stakes of these exams by allowing a retake, that women perform as well as men. This suggests that the stakes are influencing the outcome of exams so that the exams are inaccurate measures of differences between men and women. Instructors who want measurements of physics understanding less influenced by stakes (or anxiety) should adopt assessments with lower stakes.

Our discovery that gender gaps are dependent on the level of stakes associated with timed exams has an extremely important implication for education researchers. It is common for researchers to use prior exam results (e.g. quantitative SAT, ACT, \& Advanced Placement exams) as a proxy for preparation for physics. Women's scores on these exams tend to be lower than men's \cite{young_sex_2000} (This was some of the impetus for proposing research question \#2.) Therefore, a common resulting finding from those studies is that women's past preparation for physics is somehow less than men's and that this is what is driving the gender gap. An alternative interpretation of the same numerical results is that women are experiencing the same issue in both contexts, and that high-stakes exams are causing the gender gap, instead of measuring it. Therefore, this study adds evidence to our previous work indicating that it may not make sense to use exam scores as a proxy for preparation when comparing across group demographics  \cite{webb_attributing_2023}. We suggest that future research using exam scores as a proxy for preparation proceed with thoughtful caution.

\section{Limitations}

All of the retake exam classes (as well as the non-retake classes) were offered within the CLASP (Collaborative Learning through Active Sense-making in Physics) series of courses \cite{Potter2014Sixteen, paul_examining_2025}. These courses are studio type courses where students spend almost 80\% of their class time working in discussion/lab sections, generally in groups of five. The effects of allowing retake exams may well be different in classes that are predominantly lecture-based. In addition, almost the entire course grade in the courses studied was determined by the exam grades in these courses. It could be that in a course where most of the grade is determined by timed in-person solo assessments, differences of the impact of the stakes of each of those assessments are easier to identify. The data collected in this study are at least a decade old. Since that time, there have been significant changes to the educational landscape including advances in AI, a worldwide pandemic, and a changing student population. These and other factors may affect future results. 
Future research could compare exams from courses with large and small (or reverse) gender gaps to further understand how points are allocated on exams with different stakes. We encourage replication studies of the exam retake and stakes phenomena.

\section{\label{sec:Conclusions}Conclusions}

Building from our prior research on this dataset \cite{webb_attributing_2023}, where we found that retake exams were associated with a zero gender gap in introductory courses, we rule out the possibility that this happens because women are more likely to take advantage of the retake opportunity, as men are just as likely to take advantage of the retakes. We also find that the reason for the elimination of the gender gap doesn't appear to be that women need an extra chance to fail and catch up via practice or extra help (presumably due to any differences in incoming physics preparation) because women appear to do as well as men on their (low-stakes) first exam attempt in the retake classes.  These negative results are already important findings because they provide counter evidence to a recurring theme in the literature that gender gaps are due to differences in preparation.  The women in the retake classes did not need extra time nor extra tries to perform equally to the men in those classes and so showed no obvious deficiencies.

Instead we find that women do as well as and perhaps slightly better (though not significantly so) than men on the first, low stakes, exam in a retake class.  We also find that women do noticeably better overall on each of the exams (independently true after both first and second tries) in the retake course than they do in the courses without retake exams. 

Our data on final exams provide a second, and independent, measurement of an exam gender gap that is inflated by the stakes of the exam.  If the stakes of the final exam didn't matter for the gender gap, we would find that controlling for in-term exam scores (assessments of the same type, but not the same stakes) would zero the gender gap on the final exams.  Instead we find that controlling for in-term exam scores on the final exams in courses with or without retakes, does not remove the gender gap in final exams, again suggesting that the stakes of the exam control the size of the gender gap in introductory physics exams.  The size of our data set allows us to measure the extra gender gap on the final exam in these classes to better than one part in ten.

Instructors interested in removing gender performance differences related to exam stakes should reduce the stakes of their assessments by using lower stakes assessment strategies such as implementing retakes on exams.

\section{Acknowledgments}

The authors would like to thank the National Science Foundation for their partial support of this project through the now terminated NSF HSI IUSE 1953760 grant. This work would not have been possible without the support of San José State University's Research, Scholarship, and Creative Activity Assigned Time Program.

\appendix

\section{\label{sec:RetakeExamResults}Detailed retake exam data}

\begin{figure} [htb]
\includegraphics[trim=3.3cm 2.5cm 4.5cm 3.7cm, clip=true,width=\linewidth]{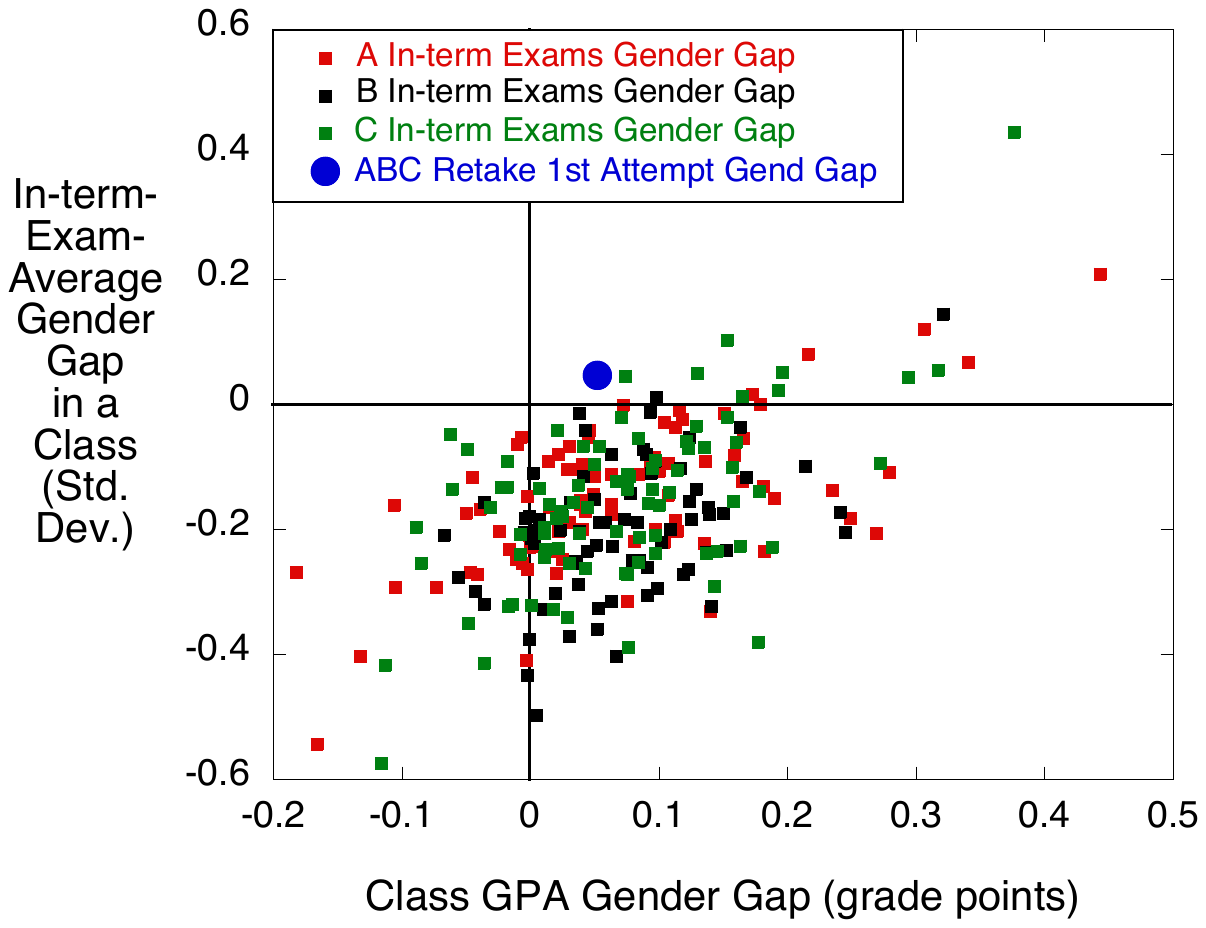}
\caption{In-term-exam average z-score for female students minus the same for male students plotted as a function of average GPA of female students minus the same for male students for 257 classes in the intro-physics for bio-scientists series at one R1 university.  The three retake classes for which we have data on the first-attempt exam scores were computed together to show the average effect upon taking an exam for the first time when a retake is possible.  No retakes were possible in the other classes so that those exams had higher stakes.}
\label{fig:AvgQvsAvgGPA}
\end{figure}

In the text we have noted that the class gender gap for in-term exams has a clear dependence on the GPA gender gap in that class. This is shown in Fig. \ref{fig:AvgQvsAvgGPA} where the class-average in-term exam gender gap is plotted as a function of that class' GPA gender gap for all 254 regular classes.  We have also grouped the first-attempts of the three retake classes together and plotted them as a single point so that one can see how an average retake class might compare to the non-retake classes.  In previous work we have noted that there seems to be a large exam/grading noise that is not easily explained by differences in student GPAs.  Thus, class-dependent gender gaps, which involve differences in noisy grade distributions might be expected to be even noisier.  The standard deviation, over classes, of the gender gap is about half of the average gender gap, showing the large variety of results even in classes that have the same instructional materials.  Nevertheless, the average retake class sits at the upper edge of the distribution over classes.

\section{\label{sec:RandStudInclusion}Including a student only once doesn't change our numbers much}

As we point out in Sec. \ref{sec:Data}, many of the students in our dataset are represented more than once because they were enrolled in, and received grades in, more than one class included in the data.  One worry we had was that including these students more than once might influence our numbers or our conclusions.  To check this we have, for each of the 26,783 unique students in the datafile, randomly chosen only one of their classes to include in an analysis.  We made 10 of these random choices in order to check the coefficients and errors in Table \ref{tab:InTermExamFit}.  The coefficient resulting from fits with those ten choices were then averaged and are shown in Table \ref{tab:RandStuChoiceFit}.  We find that the model coefficients we present in the text do not change significantly and that the error estimates change as $\sqrt{1/N}$ , where $N$ is the total number of sets of student grades, as one expects if one simply had fewer data points.  So the results and conclusions we report in the text are essentially unchanged by having a set of students present in the data more than once because they took more than one class in the series.

Our conclusions from the numerical analysis of final exam gender gaps given in Section \ref{sec:FinalExams} are also unchanged by including each student only once.  When we randomly choose only one appearance for each student in this final exam analysis we find that the gender gap is that same as shown in Table \ref{tab:FinalExamFit} but that the error estimate on the gender gap increases to 0.011 from 0.007 (so z-statistic to -10.9 from -16.2).

\begin{table}[htbp]
\caption{The average coefficients from 10 HLM fits to equation \ref{eqn:QGenderGapModel} (where each fit corresponds to a different random choice of which one class, for each student, is included) are shown along with their standard errors, z-statistics, and p-values.  The comparison with Table \ref{tab:InTermExamFit}, which includes all classes of all students, shows that the only important effect is to increase the error estimates by the square root of ratio of number of students included in the fit.  For this table the average number of students included (because we know their GPA's) is 24,641 and the retake classes include only 368 students instead of the 57,806 students and 396 students, respectively, included in Table \ref{tab:InTermExamFit}.}
\label{tab:RandStuChoiceFit}
\begin{ruledtabular}
\begin{tabular}{c c c c c}
\textbf{Coeff.} & \textbf{Value} &\textbf{Error} & \textbf{z-statistic}
& \textbf{p-value}\\ 
\hline
$b_{GPA}$ & 0.8052 & 0.0072 & 112.0 & $<10^{-3}$ \\
$b_{Female}$ & -0.2226 & 0.0072 & -31.0 & $<10^{-3}$ \\
$b_{Retake}$ & -0.164 & 0.055 & -3.00 & 0.003 \\
$b_{Female*Retake}$ & 0.233 & 0.061 & 3.85 & $<10^{-3}$ \\
\end{tabular}
\end{ruledtabular}
\end{table}

\section{\label{sec:MoreEthnicities}Results from the intersections of gender and race/ethnicity}

In our original paper \cite{webb_attributing_2023} we pointed out that the classes offering retake exams did not have statistically significantly smaller grade gaps between racial/ethnic groups historically marginalized within physics and their peers from groups that have not been marginalized.  In the interests of completeness in this paper we address the intersection of gender with race/ethnicity.  We again use Eq. \ref{eqn:QGenderGapModel} to find the GPA-controlled gender gap difference between the retake and non-retake classes but now we show the results of 12 separate fits to that model, one for each of the racial/ethnic groups in our database for which both genders are populated in both non-retake and retake classes.

\begin{figure} [htbp]
\includegraphics[trim=3.0cm 3.2cm 4.9cm 3.6cm, clip=true,width=\linewidth]{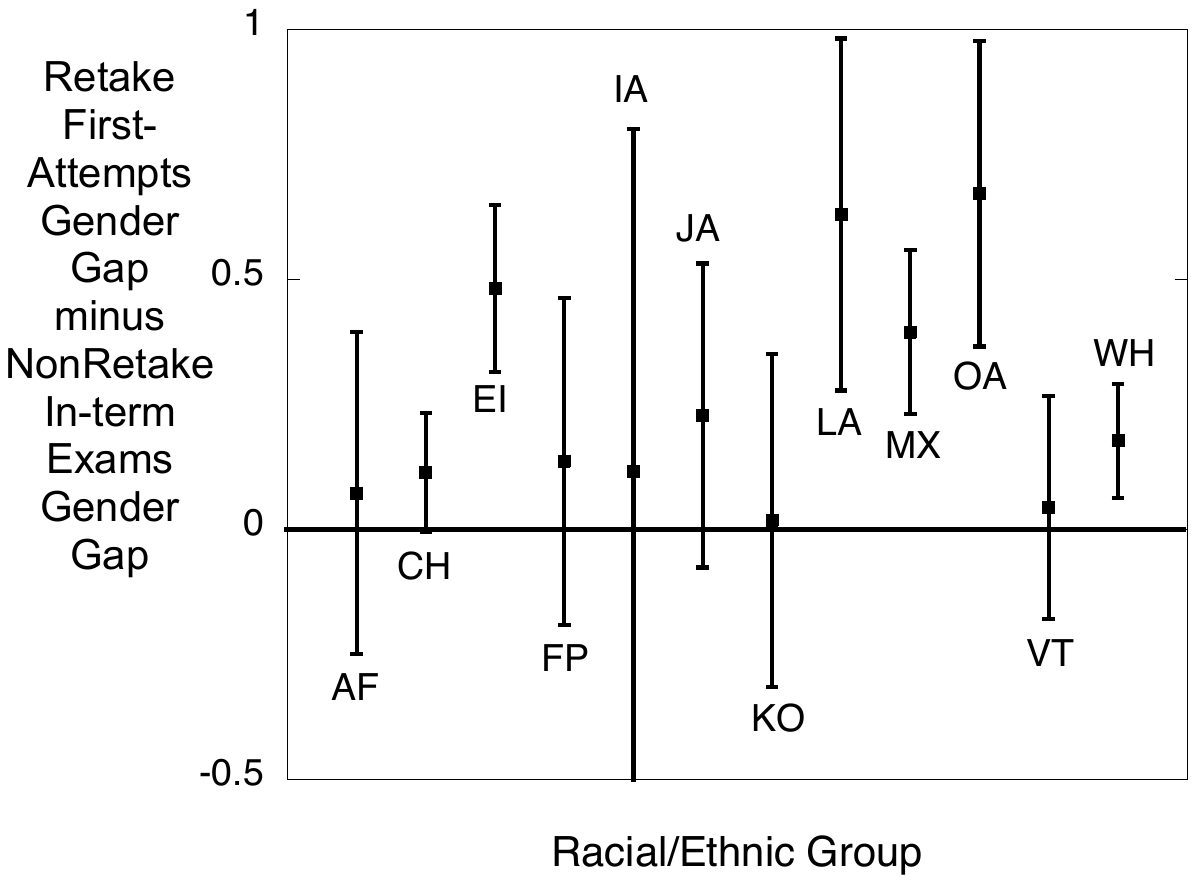}
\caption{The, GPA-controlled, gender gap differences between the first-attempt in-term exam average in the retake classes and in-term exams average in the non-retake classes.  These are shown for each separate racial/ethnic group for which there are both male students and female students identifying with that racial/ethnic group in both the retake classes and the non-retake classes.  We are actually plotting $b_{Female*Retake}$, as determined from a fit to Eq. \ref{eqn:QGenderGapModel} for each fit to the particular racial/ethnic group described in Table \ref{tab:tabA1}.}
\label{fig:Race/Ethn}
\end{figure}

In Fig. \ref{fig:Race/Ethn} we show the results for $b_{Female*Retake}$, as determined from a fit to Eq. \ref{eqn:QGenderGapModel} for each particular racial/ethnic group described in Table \ref{tab:tabA1}.  One sees that these differences are all positive although most of them are not significantly different than zero.  However, it is notable that several are well above zero.

\begin{table}[htb]
\caption{The ethnicities in Fig. \ref{fig:Race/Ethn}, as described in the data obtained from the administration of the R1 university of this paper.}
\label{tab:tabA1}
\begin{ruledtabular}
\begin{tabular}{c c}
\textbf{Symbol} & \textbf{Ethnicity} \\ 
\hline
AF & African-American/Black\\
CH & Chinese-American/Chinese\\
EI & East Indian/Pakistani \\
FP & Filipino/Filipino-American \\
IA & Indigenous American/American Indian/ \\
& Native American \\
JA  & Japanese-American/Japanese 
\\
KO & Korean-American/Korean \\
LA & Latino/Other Spanish \\
MX & Mexican-American/Mexican/ \\
& Chicano\\
OA & Other Asian-American/Other Asian \\
VT & Vietnamese-American/Vietnamese \\
WH & White/Caucasian \\
\end{tabular}
\end{ruledtabular}
\end{table}

\section{\label{sec:CoarseGrading}Grade scale effect on gender gap}

We haven't discussed issues around grade scale and gender in earlier papers so we will do some of that here.  The retake course exams discussed in this paper were graded on a very coarse grade scale.  Specifically, an answer that had the physics explained (including setting up equations) fully and correctly received one point (minor arithmetic errors were ignored), an answer that had the physics almost, but not perfectly, complete and correct received 2/3 of a point, and every other answer received zero points.  The one-point answers would have received A+ or A or A- under the regular grading in the non-retake classes and the 2/3-point answers would have received B+ or B or (maybe) B- in the non-retake classes.  One might wonder if this kind of coarse grade scale used in the retake classes could explain the lack of a gender gap in the retake classes.

Even though the regular non-retake classes used a variety of grade scales, most of these classes used either a 4-point scale or a percent scale so, for these classes, we can identify the fraction of A-grades or B-grades on specific in-term-exam items received by female students and compare to the fraction received by male students.  For this subset of 156 non-retake classes, we find that female students received A-grades on 40.02\% $\pm$ 1.3\% of their individual exam answers and B-grades on 15.24\% $\pm$ 0.07\%.  On these same exams male students received A-grades on 45.05\% $\pm$ 1.7\% of their exam answers and B-grades on 15.03\% $\pm$ 0.07\%.  So female students averaged about 5\% fewer A's and about the same number of B's as male students.  If we take these grades from the non-retake classes and rescore all A-grades to be one point, all the B-grades to be 0.67 points, and the rest of the grades to be zero, then we find the standard deviation of the resulting re-scored in-term exam grade distribution is about 0.2.  So the average normalized gender gap under this kind of coarse grading would be about (average difference of 0.05 fewer A's) multiplied by (1 point per A) and divided by (standard deviation of 0.2) = 0.25 which, given the crudeness of these grade changes, is reasonably the same as we saw for $b_{Female}$ in Table \ref{tab:InTermExamFit}.  We suggest that the coarse grading had a negligible effect on the gender gap.

\section{\label{sec:RetakeScores}Retake exams gender gap}

The group of students taking a retake exam are a self-chosen subset of the full set of students and that subset changes from exam to exam with the exam averages depending on who decides to retake an exam.  For these reasons, it is probably unwise to draw any conclusions from analyses of these data.  Nevertheless, for completeness we will calculate the gender gap on the retake exams which we have considered to be medium stakes exams.  Since the group of students retaking an exam changes from week to week we will just find the gender gap for each week's retake exam and then take a weighted average with the weights determined by the square of the standard error of that exam's gender gap.  We, again, z-score each retake exam in each class and use HLM to determine the gender gap after controlling for GPA as in Eq. \ref{eqn:IndivQGenderGapModel}.  We find that the weighted average of those four retake exam gender gaps is -0.06 $\pm$ 0.08.  This number is certainly consistent with zero gap but the 95\% confidence interval also overlaps with the other medium stakes gaps shown in Table \ref{tab:InTermExamFit}.  It probably isn't surprising that we have no conclusions to draw from these calculations.

\section{\label{sec:Nonbinary}Students who are Non-binary or for whom we don't have gender data}

Since the university didn't collect gender self-identity data until after essentially all of the exams in our dataset had been taken there is very little we can say regarding students who don't identify as one of the two genders that most populate our data.  Nevertheless, there were 5 self-identified non-binary students in our data as well as well as 31 students whose genders unknown by the university.  None of the nonbinary students appeared in a retake class.  For completeness we will extend the final exam analysis we did in Section \ref{sec:FinalExams} to include both non-binary students and students whose gender was unknown to the university.  We define a categorical variable \textit{Gender} = 0 for students listed as male in our dataset, =1 for those listed as female, =2 for those listed as non-binary, and =3 for those whose gender is unknown.  We extend our final exam analysis to include \textit{Gender} by using HLM to fit
\begin{equation}
\begin{split}
FinalExam & = b_0 + b_{InTrmExm}InTrmExm \\ &+ b_{Gender}Gender 
\end{split}
\label{eqn:FinalModelWNonBinary}
\end{equation}
where the coefficients of $Gender$ each represent the average in-term-exam-controlled final exam difference between that group and the group of male students.  All of the coefficients from the HLM fit to Eq. \ref{eqn:FinalModelWNonBinary} are given in Table \ref{tab:FinalFitNonBinary}.  The large p-values for the Non-Binary and Unknown genders suggest that, as expected, there aren't enough data to measure the differences from the male students on final exam scores.

\begin{table}[htbp]
\caption{The coefficients from an HLM fit to final exam z-scores (equation \ref{eqn:FinalModelWNonBinary}) are shown along with their standard errors, z-statistics, and p-values.  Included are 203 classes that had a total of 48,478 students. The coefficients corresponding to the different genders each represent the average difference from the group of male students.}
\label{tab:FinalFitNonBinary}
\begin{ruledtabular}
\begin{tabular}{c c c c c}
\textbf{Coeff.} & \textbf{Value} &\textbf{Error} & \textbf{z-statistic}
& \textbf{p-value}\\ 
\hline
$b_{InTrmExm}$ & 0.9652 & 0.0051 & 187.4 & $<10^{-3}$ \\
$b_{Female}$ & -0.1157 & 0.0071 & -16.3 & $<10^{-3}$ \\
$b_{NonBinary}$ & 0.27 & 0.28 & 0.96 & 034 \\
$b_{Unknown}$ & -0.06 & 0.12 & -0.50 & 0.62 \\
$b_{0}$ & 0.0744 & 0.0056 & 13.3 & $<10^{-3}$ \\
\end{tabular}
\end{ruledtabular}
\end{table}

\bibliography{references.bib}

\end{document}